\documentclass[11pt]{article}%
\usepackage{moriond,epsfig}%

\def\be{\begin{equation}}
\def\ee{\end{equation}}
\def\bea{\begin{eqnarray}}
\def\eea{\end{eqnarray}}

\begin{document}
\vspace*{4cm}
\title{CHIRAL SOLITON MODEL PREDICTIONS FOR PENTAQUARKS}

\author{M. PRASZA{\L}OWICZ }

\address{M. Smoluchowski Institute of Physics, Jagellonian Unversity, \\
ul. Reymonta 4, 30-59 Krakow, Poland}

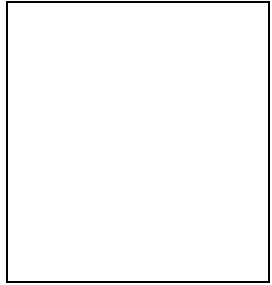
\maketitle\abstracts{We briefly describe chiral soliton model
description of baryons and predictions for exotic antidecuplet. We
discuss successful phenomenology which triggered experimental
searches and problems which arise in the formal limit of large
$N_c$.}

\section{Do we see $\Theta^{+}$ at all?}

After almost two year excitement that the exotic antidecuplet has
been discovered \cite{exp} the results from high statistics G11
experiment at CLAS were presented in April at the APS meeting with
negative result for the photoproduction of $\mathit{\Theta}^{+}$
on proton \cite{G11}. The sighting of the heaviest members of
$\overline{10}$ that were seen only by NA49 experiment at CERN
\cite{Xi} is even more problematic. Nevertheless the positive
evidence of 11 experiments that reported the existence of
$\Theta^+$ cannot be simply ignored. The reasons why some
experiments see $\Theta^+$ while the others do not maybe either of
experimental nature or a peculiar production mechanism or both.
Therefore the present confusion concerning exotics calls for a new
high precision $KN$ experiment in the interesting energy range.

\section{Chiral models}
Light antidecuplet was predicted within the chiral soliton models
($\chi$SM)~\cite{antidec}\nocite{BieDotha,Mogil,DPP}$^-$\cite{Weigel}.
Early estimate
%of the antidecuplet octet splitting,
$\mathit{\Delta} M_{\overline{10}-8}\sim 600$~MeV was obtained
already in 1984 in a specific modification of the Skyrme model
\cite{BieDotha}. The estimates of \emph{both}
$\mathit{\Theta}^{+}$ and $\mathit{\Xi}_{\overline{10}}$ masses
obtained in the Skyrme model in 1987 are in a surprising agreement
with present experimental findings \cite{Mogil}.

In this Section we will demonstrate that chiral models are deeply
rooted in QCD and take into account quark degrees of freedom maybe
even in a more complete way than the quark models themselves. The
low energy effective theory of QCD could be in principle obtained
by integrating out gluons. The resulting quark lagrangian would
preserve chiral symmetry, whose spontaneous breakdown would
produce nonzero constituent quark mass $M$ and the massless
pseudoscalar Goldstone bosons, being at the same time
$\overline{\psi} \psi$ pairs, would be present. A convenient model
of such a lagrangian is provided by a semibosonized
Nambu--Jona-Lasinio model:
\begin{equation}
\mathcal{L}=\overline{\psi}(i\not
\partial-M\,U^{\gamma_{5}}[\varphi
])\psi\label{Lagr}%
\end{equation}
which looks like a Dirac Lagrangian density for a massive fermion
$\psi$ if not for matrix $U$. In fact $\psi$ is a $3$-vector in
flavor space and also in color. Matrix
\be
U^{\gamma_{5}}=\exp\{\frac{i}{F_{\varphi}}\vec{\lambda}\cdot\vec{\varphi}%
\,\gamma_{5}\}%
\ee parameterized by a set of eight pseudoscalar fields
$\vec{\varphi}$ guaranties chiral symmetry of $\mathcal{L}$, given
by a global
multiplication of the fermion field by a phase factor%
\begin{equation}
\psi\rightarrow
e^{i\vec{\lambda}\cdot\vec{\alpha}\,\gamma_{5}}\psi
\label{chitrans}%
\end{equation}
provided we also transform meson fields%
\begin{equation}
U^{\gamma_{5}}[\varphi]\rightarrow e^{-i\vec{\lambda}\cdot\vec{\alpha}%
\,\gamma_{5}}U^{\gamma_{5}}[\varphi]\,e^{-i\vec{\lambda}\cdot\vec{\alpha
}\,\gamma_{5}}.
\end{equation}
Note that the color indices produce simply an overall factor
$N_{c}$ in front of (\ref{Lagr}).

Since the vacuum state corresponds to $U^{\gamma_{5}}=1$,
spontaneous chiral symmetry breaking indeed takes place. Moreover
the massless Goldstone bosons appear when we integrate out the
quark fields. Then the resulting effective action contains only
meson fields and can be organized in terms of a derivative expansion%
\begin{equation}
S_{\rm eff}[\varphi]=\frac{F_{\varphi}^{2}}{4}\int{\rm Tr}\left(
\partial_{\mu}U\,\partial^{\mu}U^{\dagger}\right)  +\frac{1}{32e^{2}}%
\int{\rm Tr}\left(  \left[  \partial_{\mu}U\,U^{\dagger}%
,\partial_{\nu}UU^{\dagger}\right]  ^{2}\right)
+\mathit{\Gamma}_{\rm WZ}+\ldots
\label{SkS}%
\end{equation}
where constants $F_{\varphi}$ and $e$ can be calculated from
(\ref{Lagr}) with an appropriate cut-off. $\mathit{\Gamma}_{\rm
WZ}$ is the Witten Wess-Zumino term which takes into account axial
anomaly. Perhaps the most important part are the ellipses which
encode an infinite set of terms that are effectively summed up by
the fermionic model of Eq.(\ref{Lagr}). The truncated series of
Eq.(\ref{SkS}) is the basis of the Skyrme model. Hence the Skyrme
model is (a somewhat arbitrary, because it does not include
another possible 4 derivative term) approximation to (\ref{Lagr}).

At this point both models, chiral quark model of Eq.(\ref{Lagr})
and Skyrme model of Eq.(\ref{SkS}) (without the "dots"), look like
mesonic theories describing only meson-meson
scattering~\cite{scattering}. Baryons are introduced in two steps,
following large $N_{c}$ strategy described by
Witten~\cite{Witten}. First, one constructs a soliton solution,
\emph{i.e.} solution to the classical equations of motion that
corresponds to matrix $U_0$ which cannot be expanded in a power
series around unity. Second, since the classical soliton has no
quantum numbers (except baryon number), one has to quantize the
system. Perhaps this quantization procedure, which reduces both
models to the nonrelativistic quantum system analogous to the
symmetric top \cite{antidec} with two moments of inertia
$I_{1,2}$, makes chiral-soliton models look odd and
counterintuitive.

In chiral quark soliton models stabilization of the soliton occurs
due to the valence quark level which also provides the baryon
number. In the Skyrme model where no quarks are present the
soliton is stable due to the specific choice of the 4-derivative
term in (\ref{SkS}) and the baryon number is given as a charge of
the conserved topological current. The quantization on the other
hand proceeds in both models almost identically \cite{antidec},
the only difference being that some model parameters dominated by
the valence level in the quark soliton model are exactly zero in
the Skyrme model.

\section{Exotics in chiral models}

Chiral soliton models predict that positive parity baryons fall
into SU(3) representations that contain hypercharge  $Y=N_c/3$
which is 1 in the real world. Therefore the lowest lying multiples
are octet and decuplet, exactly as in the quark models. Moreover
chiral models predict a tower of exotic rotational states starting
with $\overline {10}_{1/2}$, $27_{3/2,1/2}$,
$\overline{35}_{5/2,3/2}$ (subscripts refer to spin) etc.  The
splittings between the centers of the lowest-lying octet, decuplet
and antidecuplet baryons are given in the $\chi$SM by
\begin{equation}
\Delta M_{{10} - 8} \; = \; {3 / (2 I_1)}, \; \; \Delta
M_{{\overline {10}} - 8} \; = \; {N_c / (2 I_2)} \; = \; {3 / (2
I_2)} \label{DeltaM}
\end{equation}
where $I_{1,2}$ are two soliton moments of inertia that depend on
details of the chiral Lagrangian. Since $I_1$, $I_2\sim {\cal
O}(N_c)$, this means that $\Delta M_{{\overline {10}} - 8} \sim
{\cal O}(N_c^0)$, whereas $\Delta M_{{10} - 8}$ is  ${\cal
O}(1/N_c)$. This has triggered some
arguments~\cite{cohenlargenc,pobylitsalargenc} and
counter-arguments~\cite{DP9}, regarding the applicability of
collective coordinate quantization to the $\overline {10}$.

We have already mentioned early estimates of the antidecuplet mass
that have been recently reviewed in \cite{EKP}. The bottom line is
that antidecuplet is much lighter than in the quark models.
Therefore $\chi$SM's predict light exotic baryons belonging to
antideucuplet of positive parity.

Perhaps the most striking prediction of $\chi$SM is the small
width of the antidecuplet states. The decay width is calculated by
means of the formula for the decay width for
$B\rightarrow B^{\prime}+\varphi$:%
\begin{equation}
\mathit{\Gamma}_{B\rightarrow B^{\prime}+\varphi}=\frac{1}{8\pi}%
\frac{p_{\varphi}}{M\,M^{\prime}}\overline{\mathcal{M}^{2}}=\frac{1}{8\pi
}\frac{p_{\varphi}^{3}}{M\,M^{\prime}}\overline{\mathcal{A}^{2}}%
\label{Gammadef}%
\end{equation}
up to linear order in $m_{s}$. The \textquotedblleft bar\textquotedblright%
over the amplitude squared denotes averaging over initial and
summing over final spin (and, if explicitly indicated, over
isospin). Anticipating linear momentum dependence of the decay
amplitude $\mathcal{M}$ we have introduced reduced amplitude
$\mathcal{A}$ which does not depend on the meson momentum
$p_{\varphi}$.

Soliton models can be used to calculate the matrix element
$\mathcal{M}$. Explicitly%
\[
\mathit{\Gamma}_{B\rightarrow B^{\prime}+\varphi}=\frac{3\, G_{\mathcal{R}}^{2}%
}{8\pi M\,_{B}M_{B^{\prime}}}C_{B\rightarrow B^{\prime}+\varphi}^{\mathcal{R}%
}\,p_{\varphi}^{3}.
\]
For antidecuplet decays ($\mathcal{R}=\overline{10}$):%
\begin{equation}
G_{\overline{10}}\;=G_{0}-G_{1}-{{1}/{2}} \;G_{2},\quad
C_{\mathit{\Theta} ^{+}\rightarrow N+K}^{\overline{10}}={1}/{5},
\end{equation}
whereas for decuplet ($\mathcal{R}=10$):%
\begin{equation}
G_{10}\;=G_{0}+{{1}/{2}}\;G_{2},\quad C_{\mathit{\Delta}\rightarrow N+\pi}^{10}%
={1}/{5}.
\end{equation}
In the nonrelativistic small soliton limit \cite{limit} in which
chiral quark soliton model reproduces many results of the
nonrelativistic quark model $G_{1}/G_{0}=4/5$, $G_{2}/G_{0}=2/5$
and $G_{\overline{10}}\equiv0$! This nonintuitive cancellation
\cite{DPP} explains the small width of antidecuplet as compared to
the one of $10$ for example.

One problem concerning this cancellation is that formally%
\begin{equation}
G_{0}\sim\mathcal{O}(N_{c}^{3/2})+\mathcal{O}(N_{c}^{1/2}),\quad G_{1,2}%
\sim\mathcal{O}(N_{c}^{1/2})
\end{equation}
and it looks as if the cancellation were accidental as it occurs
between terms of different order in $N_{c}$. For arbitrary $N_{c}$
antidecuplet $\overline{10}=(0,3)$ generalizes to
$"\overline{10}"=(0,\frac{N_{c}+3}{2})$, decuplet
$"10"=(3,\frac{N_{c}-3}{2})$ and octet $"8"=(1,\frac{N_{c}-1}{2})$
\cite{largereps}, and the pertinent Clebsch-Gordan
coefficients in fact depend on $N_{c}$:%
\begin{equation}
G_{"\overline{10}"}=G_{0}-{(N_{c}+1})/{4}\; G_{1}-{1}/{2}\; G_{2}.
\end{equation}
So the subleading $G_{1}-$term is enhanced by additional factor of
$N_{c}$ and the cancellation is consistent with $N_{c}$
counting~\cite{MPGamma}.

Unfortunately there is another problem concerning the $N_c$
counting of the decay width. Because of (\ref{DeltaM})%
\begin{equation}
p_{\pi}\sim\mathcal{O}({1}/{N_{c}}),\quad p_{K}\sim\mathcal{O}%
(1)\label{pscaling}%
\end{equation}
and consequently%
\begin{equation}
\mathit{\Gamma}_{\mathit{\Delta}\rightarrow N+\pi}\sim
\mathcal{O}({1}/{N_{c}^{2}}%
),\quad\mathit{\Gamma}_{\mathit{\Theta}^{+}\rightarrow
N+K}\sim\mathcal{O}(1)
\end{equation}
in the chiral limit. This $N_{c}$ counting contradicts
experimental findings which suggest
$\mathit{\Gamma}_{\mathit{\Theta}^{+}\rightarrow
N+K}\ll\mathit{\Gamma}_{\mathit{\Delta}\rightarrow N+\pi}$.

\section{Closing remarks}

Let us finish by summarizing and by adding some remarks.

Experimental situation concerning the existence of exotic baryons
is unclear. The new data on photoproduction on deuteron from LEPS
with positive evidence have been presented on various conferences
but not published. Soon the similar data from G10 experiment at
JLab will be released, however the decisive experiment would be
certainly -- if ever completed -- high resolution KN scattering
experiment.

Little is known about the production mechanism of exotics.
Ironically this is an important factor in understanding present
experimental situation.

Most models agree that spin of antidecuplet is 1/2 in agreement
with $\chi$SM prediction. On the contrary parity is a
distinguishing feature. Were the parity of $\Theta^+$ positive as
in the $\chi$SM some other models and some lattice calculations
would require revision.

The smallness of the width is very unnatural, although $\chi$SM
provides formal explanation. If the primary decay coupling of
$\overline{10} \rightarrow 8$ is indeed very small, the SU(3)
relations between the decay rates of different members of
antidecuplet will not hold due to the flavor representation
mixing.

Other members of antidecuplet should be found. This concerns not
only $\Xi_{\overline{10}}$ but also five quark cryptoexotic
$\Sigma-$ and N$-$like states. Recent data on phototoexcitation of
nucleon resonances from GRAAl~\cite{Kuznetsov} may be interpreted
as a new narrow antidecuplet N$^*$ resonance at 1680 MeV. GRAAl
sees resonant structure only on neutron but not on proton. This
can be understood in terms of magnetic transition
moment~\cite{YangKim} $\mu_{8\rightarrow\overline{10}}$ which is
proportional to $Q-1$. Similarly modified PWA~\cite{Arndt} of
$\pi$N scattering indicates that such a state migth exist, also
STAR data show some structure in the same energy range.

There is no strong theoretical argument against pentaquarks except
its unnaturally small width. But -- as recent plethora of
theoretical papers shows -- theoretical explanation may be found
in many different models. So if high precision experiments will
not find $\Theta^+$ and its partners, this may be even more
difficult to understand than the small widths and the small mass.

\section*{Acknowledgments}

The author would like to thank the organizers of the Rencontres
de Moriond for support. The present work was also partially supported
by the Polish State Committee for Scientific Research (KBN) under
grant 2 P03B 043 24.

\end{document}